\documentclass[12pt]{article}
\usepackage{amsmath,amssymb,amsfonts,graphicx}
\newcommand{\balpha}{\mbox{\boldmath $\alpha$}}

\newcommand{\bgamma}{\mbox{\boldmath $\gamma$}}
\newcommand{\bkappa}{\mbox{\boldmath $\kappa$}}

\begin{document}
\thispagestyle{empty}
\renewcommand{\refname}{References}

%\begin{center}
\title{\bf Electronic properties of graphene with \\ a topological defect}
%\end{center}
%\begin{center}
\author{Yu.A.~Sitenko$^{1,2}$ and N.D.~Vlasii$^{1,2}$}
%\end{center}
\date{}
\maketitle
\begin{center}
$^{1}$Bogolyubov Institute for Theoretical Physics, \\
National Academy of Sciences, 03680, Kyiv, Ukraine \\
$^{2}$Physics Department, National Taras Shevchenko University of
Kyiv, \\
03127, Kyiv 127, Ukraine
\end{center}
\medskip
\medskip

Various types of topological defects in graphene are considered in
the framework of the continuum model for long-wavelength electronic
excitations, which is based on the Dirac--Weyl equation. The
condition for the electronic wave function is specified, and we show
that a topological defect can be presented as a pseudomagnetic
vortex at the apex of a graphitic nanocone; the flux of the vortex
is related to the deficit angle of the cone. The cases of all
possible types of pentagonal defects, as well as several types of
heptagonal defects (with the numbers of heptagons up to three, and
six), are analyzed. The density of states and the ground state
charge are determined.



\section{Introduction}

A synthesis of strictly twodimensional crystals composed of carbon
atoms (graphene) \cite{Nov1} is promising a wealth of new phenomena
and possible applications in technology and industry \cite{Ge}. The
observation of anomalous transport properties, and, most exciting,
the recent discovery of substantial field effect and magnetism at
room temperature allows one to envisage graphene as a reasonable
replacement of nanotubes in electronic applications, see, e.g.,
Refs. \cite{Nov2, Zh, Ne}.

By symmetry, the valence and conduction bands in graphene touch at
the corners of the hexagonal Brillouin zone. In the vicinity of
these points, the dispersion relation is isotropic and linear, and
the density of states at the Fermi level is strictly zero, rising
linearly in energy. An effective long-wavelength description of
these electronic states can be written in terms of a continuum model
which is based on the Dirac--Weyl equation for massless electrons in
2+1-dimensional space-time \cite{Di, Sem}.

Lowdimensional quantum systems of Dirac fermions can possess rather
unusual properties and, since the discovery of the effect of charge
fractionalization \cite{Ja1}, are generating current interest.
Planar Dirac fermions in the background of the
Abrikosov--Nielsen--Olesen vortex \cite{Abr, Nie} were studied in
Ref. \cite{Ja2}, and, recently, the results of this work have been
used to consider the influence of the Kekul\'{e} distortion in the
honeycomb lattice on the electronic properties of graphene
\cite{Hou, Ja3}. The present paper deals with yet another aspect,
and our purpose is to elucidate the role of topological defects in
the graphene lattice.

Topological defects appear as a result of removing (inserting) one
or several carbon atoms from (into) the honeycomb lattice without
affecting the threefold coordination of other atoms; appropriately,
the lattice  surface is warped owing to positive (negative)
curvature induced at the location of a defect. Assuming that the
size of a defect is small as compared to the whole size of the
graphene sample, our interest will be in the study of the influence
of such a defect on the electronic properties of graphene. The
consideration is based on the continuum model for long-wavelength
electronic excitations, and various types of defects are
characterized by just the number of carbon atoms removed or
inserted; actually, the size of a defect is neglected. The graphene
sheet with a defect takes shape of a cone with the value of the apex
angle related to the number of removed atoms. When a defect is
encircled, then sublattices, as well as inequivalent Fermi points,
are entwined or left untwined, depending on the type of a defect.
This imposes a condition on the electronic wave function on the
graphene sheet: a phase is acquired under a rotation around a
defect. If the phase commutes with the hamiltonian, then it can be
eliminated by a singular gauge transformation which, on the other
hand, introduces a fictitious point vortex that may be denoted in
the following as a pseudomagnetic one. The flux of the
pseudomagnetic vortex is related to the deficit (proficit) angle of
the conical surface, i.e. to the number of removed (inserted) atoms.
For certain types of defects the vortex flux takes fractional values
in the units of $2\pi$.

A general theory of planar relativistic fermionic systems in the
background of a point magnetic vortex with arbitrary flux was
proposed in Refs. \cite{Si6, Si7}; in particular, the case of
massless fermions was considered in Refs. \cite{Si9, Si0}. If the
vortex flux is fractional in the units of the London flux, then an
essentially irregular mode appears among the eigenmodes of the
one-particle hamiltonian, and this in its turn gives rize to the
appearance of an additional parameter -- the self-adjoint extension
parameter which specifies the boundary condition at the vortex point
\cite{Ger}. The theory allows us to predict the density of states
(not local but total) \cite{Si4} and the ground state quantum
numbers, as well as their local densities \cite{Si7, Si9,Si0}. The
aim of the present paper is to apply this theory to graphene with a
topological defect. The previous attempts to consider the electronic
properties of graphene with topological defects in the framework of
the continuum model approach \cite{La, Osi, Cor} have led to
contradictory and, therefore, unconvincing results. In our opinion,
this is due to the two circumstances: an inadequate choice of the
condition for the electronic wave function in the case of the
entwinement of sublattices and an inappropriate treatment of
irregular eigenmodes of the one-particle hamiltonian. These
deficiencies will be remedied in the present study.

In the next section we review the derivation of the continuum model
for planar graphene in order to specify the notations used. In
Section 3 we introduce topological defects in graphene in the
framework of the continuum model, derive the condition for the
electronic wave function in the case of an arbitrary defect, and
demonstrate that the defect can be presented as a pseudomagnetic
vortex at the apex of a graphitic nanocone. In Section 4 we consider
the solution of the Dirac--Weyl equation for electronic excitations
on a graphene sheet with a topological defect and find the density
of states and the ground state charge. The results are discussed in
Section 5. In Appendix A we show that the case of the three-pentagon
defect coincides actually with that of the absence of defects. In
Appendix B the method of a self-adjoint extension is applied to
derive the condition for the irregular solution to the Dirac--Weyl
equation.

\section{Continuum model for long-wavelength \\ electronic excitations}

Carbon atoms in graphene form a honeycomb lattice. The Bravais
lattice is triangular, and the primitive cell is rhombic and
contains two carbon atoms. If one atom is placed at the origin of
the primitive cell, another one is displaced at ${\bf d}=(-d,0)$,
where $d$ is the lattice spacing. Thus, the honeycomb lattice is
composed of two triangular sublattices: sublattice $\Lambda_A$
(black points) is generated by vectors ${\bf r}_{i}=n_i{\bf
c}_1+m_i{\bf c}_2$, and sublattice $\Lambda_B$ (blank points) is
generated by vectors ${\bf r}_{i}=n_i{\bf c}_1+m_i{\bf c}_2+{\bf
d}$, where ${\bf c}_1=\left(\frac32d,\frac{\sqrt{3}}2d\right)$ and
${\bf c}_2=\left(\frac32d,-\frac{\sqrt{3}}2d\right)$ are the basis
vectors of the primitive cell (see Fig.1a), and $n_i,\,
m_i\in\mathbb{Z}$ ($\mathbb{Z}$ is the set of integer numbers).

Each carbon atom in graphene has four valence electrons. As a result
of hybridization, three of them build $\sigma$-orbitals along the
lattice links providing for their rigidity, while the fourth one
makes $\pi$-orbital which is orthogonal to the lattice plane and is
responsible for the conductive properties of graphene. Retaining
only nearest neighbour interactions in the tight-binding
approximation, quantum-mechanical hopping of an electron on a planar
honeycomb lattice is described with the use of hamiltonian
\begin{equation}\label{1}
    {\cal H}=-t\sum\limits_{i\in \Lambda_A}\sum\limits_{j=1}^{3}a^\dagger({\bf r}_{i})b({\bf r}_{i}+{\bf
    u}_j)-t\sum\limits_{i\in \Lambda_B}\sum\limits_{j=1}^{3}b^\dagger({\bf r}_{i})
    a({\bf r}_{i}+{\bf v}_j),
\end{equation}
where $t$ is the hopping amplitude, $a^\dagger({\bf r}_{i})$ and
$a({\bf r}_{i})$ ($b^\dagger({\bf r}_{i})$ and $b({\bf r}_{i})$) are
the creation and destruction operators acting on sublattice
$\Lambda_A(\Lambda_B)$, which obey anticommutation relations
$$
[a({\bf r}_{i}),\,\,a^\dagger({\bf r}_{i'})]_+=[b({\bf
r}_{i}),\,\,b^\dagger({\bf r}_{i'})]_+=\delta_{ii'},
$$
${\bf u}_j({\bf v}_j)$ are the triad of vectors which are directed
to the nearest neighbours of an atom belonging to sublattice
$\Lambda_A(\Lambda_B)$, see Fig.1b,
\begin{equation}\label{2}
  \begin{array}{lll}
    {\bf u}_1=(-d,0), &{\bf u}_2=\left(\frac12d,\frac{\sqrt{3}}{2}d\right),
    & {\bf u}_3=\left(\frac12d,-\frac{\sqrt{3}}{2}d\right), \\
    {\bf v}_1=(d,0), & {\bf v}_2=\left(-\frac12d,-\frac{\sqrt{3}}{2}d\right),
    & {\bf v}_3=\left(-\frac12d,\frac{\sqrt{3}}{2}d\right).
  \end{array}
\end{equation}
Using the Fourier transforms of the sublattice operators,
$$
a({\bf r}_{i})=\int\limits_{\Gamma}\frac{d^2k}{(2\pi)^2}e^{i{\bf k}{\bf r}_{i}}\,\widetilde{a}({\bf k}),
\quad b({\bf r}_{i})=\int\limits_{\Gamma}\frac{d^2k}{(2\pi)^2}e^{i{\bf k}{\bf r}_{i}}\,\widetilde{b}({\bf k}),
$$
where $\Gamma$ is the first Brillouin zone, hamiltonian (1) is
presented as
\begin{equation}\label{3}
\mathcal{H}=\int\limits_{\Gamma}\frac{d^2k}{(2\pi)^2}
\,{\widetilde{\psi}}^{\,\dagger}({\bf k})\widetilde{H}\widetilde{\psi}({\bf k}),
\end{equation}
where
$$
\widetilde{\psi}({\bf k})=\left(\widetilde{a}({\bf k}),\,\widetilde{b}({\bf k})\right)^T,\quad
\widetilde{\psi}^\dagger({\bf k})=\left(\widetilde{a}^{\,\dagger}({\bf k}),\,\widetilde{b}^{\,\dagger}({\bf k})\right),
$$
and
\begin{equation}\label{3a}
    \widetilde{H}=\left(
      \begin{array}{cc}
        0 & -t\sum\limits_{j=1}^{3}e^{i{\bf k}{\bf u}_j} \\
        -t\sum\limits_{j=1}^{3}e^{i{\bf k}{\bf v}_j} & 0 \\
      \end{array}
    \right).
\end{equation}
Solving the eigenvalue problem, $\widetilde{H}\widetilde{\psi}({\bf
k})=E\widetilde{\psi}({\bf k})$, one gets
\begin{eqnarray}
  E&=&\pm t\sqrt{\sum\limits_{j=1}^{3}e^{i{\bf k}{\bf u}_j}
  \sum\limits_{j'=1}^{3}e^{i{\bf k}{\bf v}_{j'}}}=
  \nonumber \\
  &=& \pm t\sqrt{1+4\cos\left(\frac{\sqrt{3}}{2}k_yd\right)\left[\cos\left(\frac32k_xd\right)+
  \cos\left(\frac{\sqrt{3}}{2}k_yd\right)\right]}.
\end{eqnarray}
As follows from Eq.(5), the one-particle energy spectrum consists of
two surfaces ($E>0$ and $E<0$) which intersect ($E=0$) at six
conical points that are located at
\begin{equation}\label{5}
    \begin{array}{ll}
                  k_x = 0, &  k_y=\pm4\pi(3\sqrt{3}d)^{-1},\\
                  k_x=\pm2\pi(3d)^{-1}, & k_y=\pm2\pi(3\sqrt{3})d^{-1}.
                \end{array}
\end{equation}
With one electron per site, the negative-energy states (valence
band) are filled and the positive-energy states (conduction band)
are empty, so the band structure is at half-filling with the Fermi
level $E=0$ corresponding to six isolated points given in Eq.(6).
The first Brillouin zone is a hexagon with corners identified with
these Fermi points; among six of them, only two ones which can be
taken as oppositely located are inequivalent, see Fig.2.

Actually,  $\widetilde{H}$ (4) has the meaning of the one-particle
hamiltonian in the momentum representation. The low-energy
excitations can be studied by taking the continuum limit
($d\rightarrow0$) at any of two inequivalent Fermi points. Choosing
the pair of inequivalent points as ${\bf
K}_\pm=\left(0,\,\,\pm4\pi(3\sqrt{3}d)^{-1}\right)$ and keeping
terms of order ${\bf k}-{\bf K}_\pm$, one gets :
\begin{equation}\label{6}
    \widetilde{H}_\pm=\lim\limits_{d\rightarrow 0}d^{-1}\widetilde{H}|_{{\bf k}={\bf K}_\pm+{\scriptsize \bkappa}}=
\frac32t\left(\!
      \begin{array}{cc}
        0 & i\kappa_x\pm \kappa_y \\
        -i\kappa_x\pm \kappa_y & 0 \\
      \end{array}
    \!\right)=\hbar v(-\sigma^2\kappa_x\pm \sigma^1\kappa_y),
      \end{equation}
where $v=\frac 32t\hbar^{-1}$ is the Fermi velocity, and $\sigma^1$
and $\sigma^2$ are the off-diagonal Pauli matrices. Combining the
contributions from ${\bf K}_+$ and ${\bf K}_-$, one gets
\begin{equation}\label{7}
    \left(
        \begin{array}{cc}
          \widetilde{H}_+ & 0 \\
          0 & \widetilde{H}_- \\
        \end{array}
      \right)=\hbar v\left(\alpha^1\kappa_x+\alpha^2\kappa_y\right),
\end{equation}
where
\begin{equation}\label{8}
    \alpha^1=-\left(
               \begin{array}{cc}
                 \sigma^2 & 0 \\
                 0 & \sigma^2 \\
               \end{array}
             \right),\qquad \alpha^2=\left(
                                       \begin{array}{cc}
                                         \sigma^1 & 0 \\
                                         0 & -\sigma^1 \\
                                       \end{array}
                                     \right).
\end{equation}
Making the Fourier transformation of Eq.(8), one gets the
long-wavelength approximation for the one-particle hamiltonian
operator
\begin{equation}\label{9}
    H=-i\hbar v(\alpha^1\partial_x+\alpha^2\partial_y),
\end{equation}
which acts on four-component wave functions of the form
\begin{equation}\label{10}
\psi=\left(\psi_{A+},\,\psi_{B+},\,\psi_{A-},\,\psi_{B-}\right)^T,
\end{equation}
where subscripts $A$ and $B$ correspond to two sublattices and
subscripts $+$ and $-$ correspond to two inequivalent Fermi points.
Thus, an effective long-wavelength description of charge carriers in
graphene is written in terms of a continuum model which is based on
the Dirac--Weyl equation for massless electrons in $2+1$-dimensional
space-time, with the role of speed of light $c$ played by Fermi
velocity $v\approx c/300$ \cite{Di, Sem}, see also Ref. \cite{Gon}.

In order to complete the Clifford algebra of anticommuting
$4\times4$ matrices, one has to define $\gamma^0$, $\gamma^3$, and
$\gamma^5=-i\gamma^0\gamma^1\gamma^2\gamma^3$ (where
$\bgamma=\gamma^0\balpha$). It should be noted that hamiltonian (10)
commutes with generators $T_k$ ($k=1,2,3$) of the $SU(2)$-symmetry
transformations, $[T_k,\,T_l]=i\varepsilon_{klm}T_m$, where
\begin{equation}\label{11}
    T_1=\frac i2\gamma^3,\quad T_2=\frac 12\gamma^5, \quad T_3=\frac 12\gamma^3\gamma^5,
\end{equation}
and there is an arbitrariness in the choice of the representation of
the Clifford algebra, which is due to a possibility of diagonalizing
anyone of $T_k$. A representation with diagonal $\gamma^0$, in view
of the block-diagonal form of  $\alpha^1$ and $\alpha^2$, see
Eq.(9), corresponds to the choice of diagonal $T_3$ and may be
denoted as the standard planar representation:
\begin{equation}\label{12}
    \gamma^0=\left(
  \begin{array}{cc}
    \sigma^3 & 0 \\
    0 & \sigma^3 \\
  \end{array}
\right),\quad\gamma^1=i\left(
  \begin{array}{cc}
    \sigma^1 & 0 \\
    0 & \sigma^1 \\
  \end{array}
\right),\quad \gamma^2=i\left(
  \begin{array}{cc}
    \sigma^2 & 0 \\
    0 & -\sigma^2 \\
  \end{array}
\right),$$ $$\gamma^3=i\left(
  \begin{array}{cc}
    0 & \sigma^2 \\
    \sigma^2 & 0 \\
  \end{array}
\right),\quad \gamma^5=i\left(
  \begin{array}{cc}
    0 & -\sigma^2 \\
    \sigma^2 & 0 \\
  \end{array}
\right).
\end{equation}
Choosing $T_2$ to be diagonal, one gets the chiral planar
representation:
\begin{equation}\label{13}
    \gamma^0=\left(
  \begin{array}{cc}
    0 & \sigma^1 \\
    \sigma^1 & 0 \\
  \end{array}
\right),\quad\gamma^1=-i\left(
  \begin{array}{cc}
   0 & \sigma^3 \\
    \sigma^3 & 0 \\
  \end{array}
\right),\quad \gamma^2=\left(
  \begin{array}{cc}
    0 & -1 \\
    1 & 0 \\
  \end{array}
\right),$$ $$\gamma^3=-i\left(
  \begin{array}{cc}
    0 & \sigma^2 \\
    \sigma^2 & 0 \\
  \end{array}
\right),\quad \gamma^5=\left(
  \begin{array}{cc}
    -1 & 0 \\
    0 & 1 \\
  \end{array}
\right).
\end{equation}

A rotation by angle $\vartheta$ in the plane of a graphene sheet is
implemented by operator $\exp(i\vartheta\Sigma)$, where
\begin{equation}\label{14}
    \Sigma=\frac1{2i}\alpha^1\alpha^2=\frac12\left(
  \begin{array}{cc}
   \sigma^3 & 0 \\
    0 &-\sigma^3  \\
  \end{array}
\right)
\end{equation}
is the pseudospin playing here the role of the operator of spin
component which is orthogonal to the plane. The honeycomb lattice is
invariant under the rotation by $2\pi$,
\begin{equation}\label{15}
    \exp{(i2\pi\Sigma)}\,\psi=-\psi,
\end{equation}
but is not invariant under the rotation by $\pi$,
$$
\exp(i\pi\Sigma)\psi=2i\Sigma\psi,
$$
i.e. under $x\rightarrow-x$ and $y\rightarrow-y$. However, if the
latter rotation is supplemented by simultaneous exchange of
sublattices, as well as Fermi points, then this combined
transformation,
\begin{equation}\label{16}
    R\exp(i\pi\Sigma)\psi=i\left(\psi_{B-},\,\psi_{A-},\,\psi_{B+},\,\psi_{A+}\right)^T,
\end{equation}
is a symmetry one and can be regarded as the parity transformation
for graphene. Note that transformation (17) is implemented by
$i\alpha^3\equiv i\gamma^0\gamma^3$ in the standard planar
representation, see Eq.(13), or by $i\gamma^0$ in the chiral planar
representation, see Eq.(14). The explicit form of $R$ is extracted
from Eq.(17):
\begin{equation}\label{18}
    R=i\left(
  \begin{array}{cc}
    0 & \sigma^2 \\
    -\sigma^2 & 0 \\
  \end{array}
\right),
\end{equation}
and it is given by $-\gamma^5$ in representation (13) or by
$-\gamma^3\gamma^5$ in representation (14). Note that $R$ is
commuting with $\Sigma$ (15) and $H$ (10).

\section{Topological defects}

Topological defects in graphene are disclinations in the honeycomb
lattice, resulting from the substitution of a hexagon by, say, a
pentagon or a heptagon; such a disclination warps the graphene
sheet. More generally, a hexagon is substituted by a polygon with
$6-N_d$ sides, where $N_d$ is an integer which is smaller than 6.
Polygons with $N_d>0$ ($N_d<0$) induce locally positive (negative)
curvature, whereas the graphene sheet is flat away from the defect,
as is the conical surface away from the apex. In the case of
nanocones with $N_d>0$, the value of $N_d$ is related to apex angle
$\delta$,
\begin{equation}\label{18a}
\sin\frac{\delta}{2}=1-\frac {N_d}6,
\end{equation}
and $N_d$ counts the number of sectors of the value of $\pi/3$
removed from the graphene sheet, see Fig.3a. If $N_d<0$, then $-N_d$
counts the number of such sectors inserted into the graphene sheet.
Certainly, polygonal defects with $N_d>1$ and $N_d<-1$ are
mathematical abstractions, as are cones with a pointlike apex. In
reality, the defects are smoothed, and $N_d>0$ counts the number of
the pentagonal defects which are tightly clustered producing a
conical shape; such nanocones were observed experimentally
\cite{Kri}. Theory predicts also an infinite series of the
saddle-like nanocones with $-N_d$ counting the number of the
heptagonal defects clustered in their central regions. However, as
it was shown by using molecular-dynamics simulations \cite{Iha}, in
the case of $N_d\leq-4$, a surface with a polygonal defect is more
stable than a similarly shaped surface containing a multiple number
of heptagons; a screw dislocation can be presented as the
$N_d\rightarrow -\infty$ limit of a $6-N_d$-gonal defect.

The twodimensional Dirac--Weyl hamiltonian on a curved surface with
the squared length element $ds^2=g_{jj'}({\bf r})dr^jr^{j'}$ takes
form (see, e.g., Ref.\cite{Bir})
\begin{equation}\label{20}
    H=-i\hbar v{{\alpha}^{\!\!\!\!\!\sim}}^j({\bf r})\left[\partial_j+\frac i2\omega_j({\bf r})\right],
\end{equation}
where
\begin{equation}\label{21}
    \left[{{\alpha}^{\!\!\!\!\!\sim}}^j({\bf r}),\,
    {{\alpha}^{\!\!\!\!\!\sim}}^{j'}({\bf r})\right]_+= 2g^{jj'}({\bf r})I
\end{equation}
and
\begin{eqnarray}
  \omega_j({\bf r}) &=& -\frac i2{{\alpha}^{\!\!\!\!\!\sim}}^k({\bf r})
    \left[\partial_j{{\alpha}^{\!\!\!\!\!\sim}}_k({\bf r})-
    \Gamma^l_{jk}({\bf r}){{\alpha}^{\!\!\!\!\!\sim}}_l({\bf r})\right], \nonumber \\
  \Gamma^l_{jk}({\bf r})&=& \frac12g^{ln}({\bf r})\left[\partial_jg_{nk}({\bf r})+
\partial_kg_{nj}({\bf r})-\partial_ng_{jk}({\bf r})\right].
\end{eqnarray}

In the case of a conical surface with a pointlike apex, one has
\begin{equation}\label{23}
    g_{rr}=1,\quad g_{\varphi\varphi}=(1-\eta)^2r^2,
\end{equation}
where $r$ and $\varphi$ are polar coordinates centred at the apex,
and $-\infty<\eta<1$. The intrinsic curvature of a cone vanishes at
${\bf r}\neq0$ and possesses a $\delta^2({\bf r})$-singularity at
its apex (${\bf r}=0$); parameter $\eta$ enters the coefficient
before this singularity term. Introducing
$\varphi'=(1-\eta)\varphi$, one gets the metric in the
$(r,\,\varphi')$ coordinates, which is identical to that of a plane,
but with $\varphi'$ in the range $0<\varphi'<2\pi(1-\eta)$. Thus,
quantity $2\pi\eta$ for $0<\eta<1$ is the deficit angle measuring
the magnitude of the removed sector, and quantity $-2\pi\eta$ for
$-\infty<\eta<0$ is the proficit angle measuring the magnitude of
the inserted sector. In the case of graphitic nanocones, parameter
$\eta$ takes discrete values:
\begin{equation}\label{24}
    \eta=N_d/6.
\end{equation}

Using Eqs.(21) and (23), one gets
\begin{equation}\label{25}
{{\alpha}^{\!\!\!\!\!\sim}}^r=\alpha^1,\quad {{\alpha}^{\!\!\!\!\!\sim}}^\varphi=(1-\eta)^{-1}r^{-1}\alpha^2.
\end{equation}
It is straightforward to calculate the nonvanishing Christoffel
symbols
$$
\Gamma^\varphi_{r\varphi}=\Gamma^\varphi_{\varphi r}=r^{-1},\quad \Gamma^r_{\varphi\varphi}=-(1-\eta)^2r,
$$
and get the spin connection
$$
\omega_r=0,\quad \omega_\varphi=-2(1-\eta)\Sigma.
$$
Thus, hamiltonian (20) on a conical surface takes form
\begin{equation}\label{26}
    H=-i\hbar v\left\{\alpha^1\partial_r+\alpha^2r^{-1}\left[(1-\eta)^{-1}\partial_\varphi-i\Sigma\right]\right\}.
\end{equation}

In the case of the planar graphene sheet ($\eta=0$), wave function
(11) satisfyes antiperiodicity condition, see Eq.(16),
\begin{equation}\label{27}
    \psi(r,\,\varphi+2\pi)=-\psi(r,\,\varphi),
\end{equation}
i.e. the wave function is a section of a bundle with spin connection
$-2\Sigma$.

Let us consider a graphene sheet with a pentagonal disclination
($N_d=1$). When circling once this defect, the two sublattices in
the honeycomb structure are exchanged (see Fig.3b), the two
inequivalent Fermi points are exchanged as well. Circling twice this
defect is analogous to circling once a hexagon in the honeycomb
lattice without defects. The situation resembles that of a
M\"{o}bius strip, where a double full turn is needed to arrive at
the same point. Thus, in the continuum model description of graphene
with a pentagonal disclination, wave function (11) has to obey the
M\"{o}bius-strip-type condition:
\begin{equation}
  \psi(r,\,\varphi+2\pi) = \Biggl.iR \psi(r,\,\varphi),
\end{equation}
where $R$ is given by Eq.(18), and, consequently,
\begin{equation}
  \psi(r,\,\varphi+4\pi) =-\psi(r,\,\varphi),
\end{equation}
since $R^2=I$. Note that the sign in the right hand side of Eq.(28)
is chosen by convention.

In a similar way, one can show that the wave function on a graphene
sheet with a heptagonal disclination ($N_d=-1$) obeys the
M\"{o}bius-strip-type condition as well. Moreover, it can be noted
that sublattices (and Fermi points) are entwined in the case of odd
$N_d$ and are left untwined in the case of even $N_d$. Thus, the
condition for the wave function on a graphene sheet with an
arbitrary disclination takes form
\begin{equation}\label{30}
    \psi(r,\,\varphi+2\pi)=-\exp\left(-i\frac\pi2N_dR\right)\psi(r,\,\varphi),
\end{equation}
where the choice of sign in the exponential function agrees with the
sign choice in Eq.(28). Our results remain unchanged if the opposite
sign in Eq.(28) and, correspondingly, in the exponential function in
Eq.(30) is chosen.

By performing singular gauge transformation
\begin{equation}\label{31}
    \psi'=e^{i\Omega}\psi,\quad \Omega=\varphi\frac{N_d}{4}R,
\end{equation}
one gets the wave function obeying condition
\begin{equation}\label{32}
    \psi'(r,\,\varphi+2\pi)=-\psi'(r,\,\varphi),
\end{equation}
in the meantime, hamiltonian (26) is transformed to
\begin{equation}\label{33}
    \!\!\!\!H'\!=\!e^{i\Omega}He^{-i\Omega}\!=\!-i\hbar v\left\{\alpha^1\partial_r+\alpha^2r^{-1}\left[
    (1-\eta)^{-1}\!\left(\partial_\varphi-i\frac{3}{2}\eta R\right)\!-i\Sigma\right]\right\},
\end{equation}
where Eq.(24) is recalled. We conclude that a topological defect in
graphene is presented by a pseudomagnetic vortex with flux
$N_d\,\pi/2$ through the apex of a cone with deficit angle
$N_d\,\pi/3$.

\section{Ground state charge}

The density of states is defined as
\begin{equation}\label{34}
    \tau(E)=\frac1\pi {\rm Im}\,{\rm Tr}(H-E-i0)^{-1},
\end{equation}
where ${\rm Tr}$ is the trace of an integro-differential operator in
the functional space: ${\rm Tr}O=\int d^2r{\rm tr}\langle{\bf
r}|O|{\bf r}\rangle$; ${\rm tr}$ denotes the trace over spinor
indices only. Since the continuum model of graphene without
topological defects corresponds to the use of the free Dirac--Weyl
hamiltonian in flat twodimensional space, see Eq.(10), the density
of states is immediately calculable and found to be proportional to
the size of the graphene sheet
\begin{equation}\label{35}
    \tau(E)=\frac{S|E|}{\pi\hbar^2v^2},
\end{equation}
where $S$ is its area. The ground state charge of the graphene
sheet,
\begin{equation}\label{36}
    Q=-\frac e2\int\limits_{-\infty}^{\infty}dE\,\tau(E)\,{\rm sgn}(E),
\end{equation}
is evidently zero, because Eq.(35) is even in energy.

To consider the influence of topological defects in graphene on the
density of states in the framework of the continuum model, we need
the complete set of solutions to the Dirac--Weyl equation in this
case
\begin{equation}\label{37}
    H'\psi'=E\psi',
\end{equation}
where $H'$ is given by Eq.(33), and $\psi'$ is the spinor wave
function obeying condition (32). In general, the contribution of a
topological defect is added to Eq.(35), and, lacking the factor of
area, it seems to be negligible. However, if this contribution
contains a piece which is odd in energy, then the latter yields the
nonzero ground state charge, see Eq.(36). In the following our
interest will be in the search of such a piece, and, as we shall
see, its emergence is due to the appearance of an irregular solution
to the Dirac--Weyl equation.

Let us make unitary transformation
\begin{equation}\label{38}
    \psi''=U\psi',\quad H''=UH'U^{-1},
\end{equation}
where
\begin{equation}\label{39}
    U=U^{-1}=\frac1{\sqrt{2}}\left(
                               \begin{array}{cc}
                                 I & i\sigma^2 \\
                                 -i\sigma^2 & -I \\
                               \end{array}
                             \right),
\end{equation}
then
\begin{equation}\label{40}
    URU^{-1}=\left(
                              \begin{array}{cc}
                                I & 0 \\
                                0 & -I \\
                              \end{array}
                            \right)
\end{equation}
and transformed hamiltonian $H''$ acquires a block-diagonal form:
\begin{equation}\label{41}
    H''=\left(
                               \begin{array}{cc}
                                 H_1 & 0 \\
                                 0 & H_{-1} \\
                               \end{array}
                             \right),
\end{equation}
where
\begin{equation}\label{42}
    \!\!\!\!\!H_s=\hbar v\left\{i\sigma^2\partial_r-\sigma^1r^{-1}\left[(1-\eta)^{-1}\left(
    is\partial_\varphi+\frac32\eta\right)+\frac12\sigma^3\right]\right\}, \,\,\, s=\pm1.
\end{equation}
It should be emphasized that the definite sublattice ($A$ or $B$)
and Fermi-point ($+$ or $-$) indices are assigned to the components
of $\psi$ (11), while, after performing transformations (31) and
(38), one gets $\psi''$ with components mixing up different
sublattices and Fermi points. Certainly, the calculation of
functional trace in Eq.(34) does not depend on the representation
used, and it is just a matter of convenience to use a representation
with the block-diagonal form of hamiltonian (41).

Separating the radial and angular variables
\begin{equation}\label{43}
    \psi''(r,\varphi)=\sum\limits_{n\in \mathbb{Z}}\langle r,\varphi|E,n\rangle,
\end{equation}
where
\begin{equation}\label{44}
\langle r,\varphi|E,n\rangle=\left(
                               \begin{array}{c}
                                 f_{n,1}(r)\, e^{i(n+\frac12)\varphi} \\
                                 g_{n,1}(r)\, e^{i(n+\frac12)\varphi} \\
                                 f_{n,-1}(r)\, e^{i(n-\frac12)\varphi} \\
                                 g_{n,-1}(r)\, e^{i(n-\frac12)\varphi} \\
                               \end{array}
                             \right),
\end{equation}
one rewrites the Dirac--Weyl equation as the system of equations for
the radial functions
\begin{equation}\label{45}
    \left(
      \begin{array}{cc}
        0 & D^\dagger_{n,s} \\
        D_{n,s} & 0 \\
      \end{array}
    \right)\left(
             \begin{array}{c}
               f_{n,s}(r) \\
               g_{n,s}(r) \\
             \end{array}
           \right)=E\left(
                      \begin{array}{c}
                       f_{n,s}(r) \\
               g_{n,s}(r) \\
                      \end{array}
                    \right),
\end{equation}
where
\begin{eqnarray}
  D_{n,s} &=& \hbar v\left[-\partial_r+r^{-1}(1-\eta)^{-1}(sn-\eta)\right],  \nonumber \\
  D_{n,s}^\dag &=& \hbar v\left[\partial_r+r^{-1}(1-\eta)^{-1}(sn+1-2\eta)\right].
\end{eqnarray}
A pair of linearly independent solutions to Eq.(45) is written in
terms of the cylinder functions. In the case of $\frac12\leq\eta<1$
($N_d=3,\,4,\,5$) the condition of regularity at the origin is
equivalent to the condition of square integrability at this point,
and this selects a physically reasonable solution. Thus, the
situation is similar to that of $\eta=0$ (absence of a defect),
resulting in a density of states which is even in energy. In
particular, it can be shown that the density of states in the case
of $\eta=\frac12$ ($N_d=3$) is given by Eq.(35), see Appendix A.

In the case of $0<\eta<\frac12$ ($N_d=1,\,2$) and $-\frac12\leq
\eta<0$ ($N_d=-1,\,-2,\,-3$) there is a mode, for which the
condition of regularity at the origin is not equivalent to the
condition of square integrability at this point: both linearly
independent solutions for this mode are at once irregular and square
integrable at the origin. To be more precise, let us define in this
case
\begin{equation}\label{47}
    n_c=\frac s2\left[{\rm sgn}(\eta)-1\right],
\end{equation}
and note that solutions to the Dirac--Weyl equation correspond to
the continuous spectrum and, therefore, obey the orthonormality
condition
\begin{equation}\label{48}
    \int\limits_{0}^{2\pi}d\varphi \int\limits_{0}^{\infty}dr\,r(1-\eta)\langle E,n|r,\varphi\rangle
    \langle r,\varphi|E',n'\rangle=\frac{2\delta(E-E')}{\sqrt{EE'}}\delta_{nn'},
\end{equation}
where a factor of 2 in the right hand side of the last relation is
due to the existence of two inequivalent Fermi points. Then the
complete set of solutions to Eq.(45) is chosen in the following
form:
\newline regular modes with $sn>sn_c$

\begin{equation}\label{49}
    \left(
      \begin{array}{c}
        f_{n,s}(r) \\
        g_{n,s}(r) \\
      \end{array}
    \right)=\frac1{2\sqrt{\pi(1-\eta)}}\left(
                                          \begin{array}{c}
                                            J_{l(1-\eta)^{-1}-F}(kr) \\
                                            {\rm sgn}(E)J_{l(1-\eta)^{-1}+1-F}(kr) \\
                                          \end{array}
                                        \right),\quad l=s(n-n_c),
\end{equation}

\noindent regular modes with $sn<sn_c$
\begin{equation}\label{50}
    \left(
      \begin{array}{c}
        f_{n,s}(r) \\
        g_{n,s}(r) \\
      \end{array}
    \right)=\frac1{2\sqrt{\pi(1-\eta)}}\left(
                                          \begin{array}{c}
                                            J_{l'(1-\eta)^{-1}+F}(kr) \\
                                            -{\rm sgn}(E)J_{l'(1-\eta)^{-1}-1+F}(kr) \\
                                          \end{array}
                                        \right),\quad l'=s(n_c-n),
\end{equation}
and an irregular mode
\begin{eqnarray}\label{51}
 \left(
      \begin{array}{c}
        f_{n_c,s}(r) \\
        g_{n_c,s}(r) \\
      \end{array}
    \right)=\frac1{2\sqrt{\pi(1-\eta)\left[1+\sin(2\nu_E)\cos(F\pi)\right]}}\times \nonumber \\
    \times\left(
                                                                               \begin{array}{c}
                                                                                 \sin(\nu_E)J_{-F}(kr)+\cos(\nu_E)J_{F}(kr) \\
                                                                                 {\rm sgn}(E)\left[\sin(\nu_E)J_{1-F}(kr)-\cos(\nu_E)J_{-1+F}(kr)\right] \\
                                                                               \end{array}
                                                                             \right),
\end{eqnarray}
where $k=|E|(\hbar v)^{-1}$, $J_\mu(u)$ is the Bessel function of
order $\mu$, and

\begin{equation}\label{52}
    F=\left[\frac12-\frac12 {\rm sgn}(\eta)+\eta\right](1-\eta)^{-1}.
\end{equation}

Thus, the requirement of regularity for all modes is in
contradiction with the requirement of completeness for these modes.
The problem is to find a condition allowing for irregular at
$r\rightarrow 0$ behaviour of the mode with $n=n_c$, i.e. to fix
$\nu_E$ in Eq.(51). To solve this problem, first of all one has to
recall the result of Ref. \cite{Wei}, stating that for the partial
Dirac hamiltonian to be essentially self-adjoint, it is necessary
and sufficient that a non-square-integrable (at $r\rightarrow 0$)
solution exist. Since such a solution does not exist in the case of
$n=n_c$, the appropriate partial hamiltonian is not essentially
self-adjoint. The Weyl-von Neumann theory of self-adjoint operators
(see, e.g., Ref. \cite{Alb}) is to be employed in order to consider
a possibility of the self-adjoint extension for this operator. We
show in Appendix B that the self-adjoint extension exists indeed,
and the partial hamiltonian at $n=n_c$ is defined on the domain of
functions obeying the condition

\begin{equation}\label{53}
    \frac{\lim\limits_{r\rightarrow 0}(rMv/\hbar)^Ff_{n_c,s}(r)}
    {\lim\limits_{r\rightarrow 0}(rMv/\hbar)^{1-F}g_{n_c,s}(r)}=
    -2^{2F-1}\frac{\Gamma(F)}{\Gamma(1-F)}\tan\left(\frac \Theta 2+\frac\pi 4\right),
\end{equation}

\noindent where $\Gamma(u)$ is the Euler gamma function, $M$ is the
parameter of the dimension of mass, and $\Theta$ is the self-adjoint
extension parameter. Substituting the asymptotics of Eq.(51) at
$r\rightarrow 0$ (see, e.g., Ref. \cite{Abra}) into Eq.(53), one
gets the relation fixing parameter $\nu_E$,

\begin{equation}\label{54}
    \tan(\nu_E)={\rm sgn}(E)\left(\frac{\hbar k}{Mv}\right)^{2F-1}\tan\left(\frac \Theta 2+\frac\pi 4\right).
\end{equation}

Using the complete set of solutions, it is straightforward to
determine the kernel of the resolvent, $\langle
r,\,\varphi|(H-\omega)^{-1}|r',\,\varphi'\rangle$ (where $\omega$ is
a complex parameter with dimension of energy), calculate its
functional trace, see, e.g., Ref. \cite{Si4}, and find density of
states (34). Only the irregular mode contributes to the odd in
energy piece of the density of states, which is given by expression

$$
\hspace{-12.2cm}\tau(E)=
$$
\begin{equation}
    =\frac{2(2F-1)\sin(F\pi)\left[\left(\frac{|E|}{Mv^2}\right)^{2F-1}
    \tan\left(\frac \Theta 2+\frac\pi 4\right)+\left(\frac{|E|}{Mv^2}\right)^{1-2F}
    \cot\left(\frac \Theta 2+\frac\pi 4\right)\right]}{\pi E\left[\left(\frac{|E|}{Mv^2}\right)^{2(2F-1)}
    \tan^2\left(\frac \Theta 2+\frac\pi 4\right)\!-\!2\cos(2F\pi)\!+\!\left(\frac{|E|}{Mv^2}\right)^{2(1-2F)}\cot^2\left(\frac
    \Theta 2+\frac \pi 4\right)\right]}.
\end{equation}
Inserting Eq.(55) into Eq.(36), we calculate the ground state
charge,
\begin{equation}\label{56}
    Q= e\,{\rm sgn}_0[(1-2F)\cos\Theta],
\end{equation}
where
$$
{\rm sgn}_0(u)=\left\{\begin{array}{cc}
                        {\rm sgn}(u), & u\neq 0 \\
                        0, & u=0
                      \end{array}
\right\}.
$$
The charge, if any, is accumulated in the vicinity of the defect,
and its density is given by expression, see Ref. \cite{Si0},
\begin{equation}\label{57}
    \rho(r)=e\frac{2\sin(F\pi)}{\pi^3(1-\eta)r^2}\int\limits_{0}^{\infty}
    \frac{dw\,w\left[K_{1-F}(w)-K_F(w)\right]}{\left(\frac{\hbar w}{rMv}\right)^{2F-1}\tan
    \left(\frac{\Theta}{2}+\frac\pi4\right)+\left(\frac{\hbar w}{rMv}\right)^{1-2F}\cot
     \left(\frac{\Theta}{2}+\frac\pi4\right)},
\end{equation}
decreasing as an inverse power at large distances from the defect;
here $K_\mu(u)$ is the Macdonald function of order $\mu$.

In the case of $\eta<-\frac12$ ($N_d=-4,\,-5,\,\ldots$) there are
two or more irregular modes, unless $\eta=-1$ ($N_d=-6$). The case
of more than one irregular modes will be considered elsewhere, while
in the case of $\eta=-1$ the irregular mode appears at $n=n_c$ with
$n_c=-2s$ and has the form
\begin{equation}\label{58}
    \left(
      \begin{array}{c}
        f_{n_c,s}(r) \\
        g_{n_c,s}(r) \\
      \end{array}
    \right)=\frac{1}{2\sqrt{\pi(1-\eta)}}\left(
                                            \begin{array}{c}
                                              \sin(\nu_E)J_{-\frac12}(kr)+\cos(\nu_E)J_{\frac12}(kr) \\
                                              {\rm sgn}(E)\left[\sin(\nu_E)J_{\frac12}(kr)-\cos(\nu_E)J_{-\frac12}(kr)]\right] \\
                                            \end{array}
                                          \right);
\end{equation}
hence this case corresponds to $F=\frac12$ in Eq.(51), yielding the
vanishing ground state charge. Our results are summarized in the
Table.

If the sign in the exponential function in condition (30) is changed
to the opposite, then this corresponds to change $F\rightarrow 1-F$.
Our results remain unchanged, if, in addition, shift
$\Theta\rightarrow \Theta+\pi$ is performed.

\section{Discussion}

In the present paper we study the electronic properties of the
carbon monolayer (graphene) with disclinations, i.e. $6-N_d$-gonal
($N_d\neq 0$) defects inserted in the otherwise perfect
twodimensional hexagonal lattice. The effects of the variation of
the bond length or the mixing of $\pi$- with $\sigma$-orbitals
caused by curvature of the lattice surface are neglected, and our
consideration, focusing on global aspects of coordination of carbon
atoms, is based on the long-wavelength continuum model originating
in the tight-binding approximation for the nearest neighbour
interactions. Our general conclusion is that the dependence of the
electronic properties on the value of $N_d$ is not monotonic, but
rather abruptly discontinuous. For some values of $N_d$ the density
of states is predicted unambiguously by the theory, whereas,
otherwise, its theoretical prediction involves some parameters which
should be determined from the experiment.

As it was already noted \cite{La}, a defect with odd $N_d$ entwines
two sublattices, as well as two inequivalent Fermi points, and in
the present paper we show that the correct condition for the
electronic wave function involves operator $R$ commuting with the
hamiltonian, see Eqs.(30) and (18). However, much stronger impact on
electronic properties might be drawn by the fact that for certain
$N_d$ irregular modes emerge among the eigenmodes of the
hamiltonian. It is instructive to compare two cases when the density
of states remains almost the same as for planar graphene, but for
different reasons. In the case of the three-pentagon defect
($N_d=3$), there is entwinement of sublattices and there is no
irregular modes; the density of states is calculated (see Appendix
A) and shown to coincide exactly with that of planar graphene. In
the case of the two-pentagon defect ($N_d=2$), there is no
entwinement and there is an irregular mode; the density of states is
even in energy (owing to $F=\frac 12$, the odd piece vanishes, see
Eq.(55)) and almost coincides with that of planar graphene. Thus, we
disprove the controversial assertions that the density of states at
the Fermi level is nonzero either at $N_d=2$ \cite{La} or at $N_d=3$
\cite{Osi}.

The same unambiguous predictions are obtained for a graphene sheet
with the two-heptagon defect ($N_d=-2$) and a graphene sheet with a
dodecagon or six heptagons ($N_d=-6$): the density of states almost
coincides with that of planar graphene. Evidently, the ground state
charge is zero in all above cases.

Let us turn now to the cases when our predictions are not
unambiguous, since they involve self-adjoint extension parameter
$\Theta$. These cases include graphene sheets with following
defects: one pentagon ($N_d=1$), one heptagon ($N_d=-1$), and three
heptagons ($N_d=-3$). Contrary to the assertions in Refs.\cite{La,
Osi}, the density of states at the Fermi level is characterized by a
divergent, rather than the cusp, behaviour in these cases, see
Eq.(55):
\begin{equation}\label{59}
    \tau(E)    \begin{array}{c}
               = \\[-0.2cm]
               {E\rightarrow 0} \\
             \end{array}\left\{\begin{array}{cc}
                          -\frac{6}{5\pi}\frac{{\rm sgn}(E)}{Mv^2}\left(\frac{Mv^2}{|E|}\right)^{\frac 25}\cot
                          \left(\frac\Theta 2+\frac \pi 4\right), & N_d=1, \\
                          \frac{6}{7\pi}\frac{{\rm sgn}(E)}{Mv^2}\left(\frac{Mv^2}{|E|}\right)^{\frac 47}\tan
                          \left(\frac\Theta 2+\frac \pi 4\right), & N_d=-1, \\
                          -\frac{2}{3\pi}\frac{{\rm sgn}(E)}{Mv^2}\left(\frac{Mv^2}{|E|}\right)^{\frac 23}\cot
                          \left(\frac\Theta 2+\frac \pi 4\right), & N_d=-3.
                        \end{array}\right.
\end{equation}
Actually, there are three possibilities: $\cos\Theta>0$,
$\cos\Theta<0$, and $\cos\Theta=0$. The question of which of the
possibilities is realized has to be answered by experimental
measurements. First, the density of states in the vicinity of the
Fermi level can be measured directly in scanning tunnel and
transmission electron microscopy. Secondly, the ground state charge
can be measured also, and our prediction, see Eq.(56) or the Table,
is
\begin{equation}\label{59a}
Q|_{N_d=1}=-Q|_{N_d=-1}=Q|_{N_d=-3},
\end{equation}
while the ground state charge density decreases by power law at
large distances from the defect, see Eq.(57),
\begin{equation}\label{60}
    \rho(r)    \begin{array}{c}
               = \\[-0.2cm]
               {r\rightarrow \infty} \\
             \end{array}\left\{\begin{array}{cc}
                          e\frac{9\sin(\pi/5)}{20\pi^2}\frac{\Gamma\left(\frac{13}{10}\right)\Gamma\left(
                          \frac{11}{10}\right)}{\Gamma\left(\frac{4}{5}\right)}\frac{1}{r^2}\left(
                          \frac{\hbar}{rMv}\right)^{\frac35}\cot\left(\frac\Theta2+\frac\pi4\right), & N_d=1, \\
                          -e\frac{9\sin(2\pi/7)}{35\pi^2}\frac{\Gamma\left(\frac{17}{14}\right)\Gamma\left(
                          \frac{13}{14}\right)}{\Gamma\left(\frac{5}{7}\right)}\frac{1}{r^2}\left(
                          \frac{\hbar}{rMv}\right)^{\frac37}\tan\left(\frac\Theta2+\frac\pi4\right), & N_d=-1, \\
                          e\frac{\sqrt{3}}{12\pi^2}\frac{\Gamma\left(\frac{7}{6}\right)\Gamma\left(
                          \frac{5}{6}\right)}{\Gamma\left(\frac{2}{3}\right)}\frac{1}{r^2}\left(
                          \frac{\hbar}{rMv}\right)^{\frac13}\cot\left(\frac\Theta2+\frac\pi4\right), & N_d=-3.
                        \end{array}\right.
\end{equation}
The results for the ground state charge in the case of
$\cos\Theta>0$ agree with the results of numerical atomistic
calculation of the bond network with the use of recursion methods
\cite{Tam}. The pentagonal defect, as well as the three-heptagon
one, is attractive, and the heptagonal defect is repulsive to
electrons. The charge, negative or positive, is accumulated around
the defect, and, at large distances from it, the decrease is the
strongest one for a pentagon and the weakest one for three
heptagons.

It should be noted that at $\cos\Theta\neq 0$ and $F\neq \frac12$
scale invariance is broken, and the appearance of parameter $M$ with
dimension of mass evinces this. In general, irregular mode (51)
diverges at the origin as $r^{-\nu}$ with $\nu<1$. Scale invariance
is respected by the condition of minimal irregularity \cite{Si90,
Si6, Si7},
\begin{equation}\label{60a}
    \Theta=\left\{\begin{array}{cc}
             \frac\pi2({\rm mod}2\pi), & 0<F<\frac12, \\
             -\frac\pi2({\rm mod}2\pi), & \frac12<F<1,
           \end{array}\right.
\end{equation}
which restricts the behaviour of the irregular mode at the origin to
$r^{-\nu}$ with $\nu<\frac12$. Thus, both scale invariance and
minimal irregularity favour definitely the choice of $\cos\Theta=0$,
when the density of states and the ground state charge are trivial.
It would be inspiring, if the experiment could prefer other choices.

\section*{Acknowledgements}

We would like to thank V.P. Gusynin for stimulating discussions. The
research was supported in part by the Swiss National Science
Foundation under the SCOPES project No. IB7320-110848. Yu.A.S.
acknowledges the support of the State Foundation for Fundamental
Research of Ukraine (grant F16-457-2007) and INTAS (grant No.
05-1000008-7865). N.D.V. acknowledges the INTAS support through the
PhD Fellowship Grant for Young Scientists (No. 05-109-5333).

\renewcommand{\thesection}{A}
\renewcommand{\theequation}{\thesection.\arabic{equation}}
\setcounter{section}{1} \setcounter{equation}{0}

\section*{Appendix A}

In the case of $\eta=\frac12$ hamiltonian $H_s$ (42) takes form
\begin{equation}\label{a1}
    H_s=\hbar v\left[i\sigma^2\partial_r-\sigma^1r^{-1}\left(2is\partial_\varphi+\frac32
    +\frac12\sigma^3\right)\right].
\end{equation}
The kernel of the resolvent (the Green's function) of the
hamiltonian is presented as
\begin{equation}\label{a2}
    \langle r,\,\varphi|(H_s-\omega)^{-1}|r',\,\varphi'\rangle=\frac 1{2\pi}\sum\limits_{n\in \mathbb{Z}}e^{i
    \left(n+\frac s2\right)(\varphi-\varphi')}\left(
                                                \begin{array}{cc}
                                                  a^{11}_n(r,\,r') & a^{21}_n(r,\,r')\\
                                                  a^{12}_n(r,\,r') & a^{22}_n(r,\,r')\\
                                                \end{array}
                                              \right),
\end{equation}
where the radial components satisfy equations
$$
  \left(\!
    \begin{array}{cc}
      -\omega & \hbar v(\partial_r+r^{-1}2sn) \\
      \hbar v\left[-\partial_r+r^{-1}(2sn\!-\!1)\right] & -\omega \\
    \end{array}
 \! \right)\left(\!
                                                \begin{array}{cc}
                                                  a^{11}_n(r,\,r') & a^{21}_n(r,\,r')\\
                                                  a^{12}_n(r,\,r') & a^{22}_n(r,\,r')\\
                                                \end{array}
                                              \!\right)=
$$
$$
  \!=\!\left(\!
    \begin{array}{cc}
      -\omega & \hbar v(\partial_{r'}+{r'}^{-1}2sn) \\
      \!\!\hbar v\left[-\partial_{r'}+{r'}^{-1}(2sn\!-\!1)\right] & -\omega \\
    \end{array}
  \!\right)\left(\!
                                                \begin{array}{cc}
                                                  a^{11}_n(r,\,r') & a^{12}_n(r,\,r')\\
                                                  a^{21}_n(r,\,r') & a^{22}_n(r,\,r')\\
                                                \end{array}
                                              \!\right)\!=
$$
\begin{equation}
     =\frac 2r\delta(r-r')\left(
                         \begin{array}{cc}
                           1 & 0 \\
                           0 & 1 \\
                         \end{array}
                       \right);
\end{equation}
note that a factor before the delta-function is due to $({\rm
det}g_{jj'})^{-\frac12}=[(1-\eta)r]^{-1}$. The radial components can
be found in the form
\begin{equation}\label{a3}
    a^{kk'}_n=\int\limits_{0}^{\infty}\frac{dp\,p}{\hbar^2v^2p^2-\omega^2}a_{n,p}^{kk'}(r,r'),
\end{equation}
where
\begin{eqnarray}
  a_{n,p}^{11}(r,r') &=& 2\omega\,J_{2sn-1}(pr)J_{2sn-1}(pr'), \nonumber \\
  a_{n,p}^{12}(r,r') &=& 2\hbar v p\,J_{2sn}(pr)J_{2sn-1}(pr'),\nonumber\\
  a_{n,p}^{21}(r,r') &=& 2\hbar v p\,J_{2sn-1}(pr)J_{2sn}(pr'),\\
  a_{n,p}^{22}(r,r') &=& 2\omega\,J_{2sn}(pr)J_{2sn}(pr').\nonumber
\end{eqnarray}
Putting $\varphi'=\varphi$ and taking trace of matrix (A.2), one
gets
$$
{\rm tr}\langle r,\varphi|(H_s-\omega)^{-1}|r',\varphi\rangle=\frac1{2\pi}\sum\limits_{n\in \mathbb{Z}}\left[
a_n^{11}(r,r')+a_n^{22}(r,r')\right]=
$$
\begin{equation}\label{a4}
=\frac \omega\pi\int\limits_{0}^{\infty}\frac{dp\,p}{\hbar^2v^2p^2-\omega^2}
\sum\limits_{n\in \mathbb{Z}}\left[J_{2sn-1}(pr)J_{2sn-1}(pr')+J_{2sn}(pr)J_{2sn}(pr')\right].
\end{equation}
The summation is performed with the use of the Neumann's addition
theorem (see, e.g., Ref. \cite{Abra}), yielding the expression,
\begin{equation}\label{a5}
    {\rm tr} \langle r,\varphi|(H_s-\omega)^{-1}|r',\varphi\rangle=\frac
    \omega\pi\int\limits_{0}^{\infty}\frac{dp\,p}{\hbar^2v^2p^2-\omega^2}
J_0[p(r-r')],
\end{equation}
which is badly divergent at $r'\rightarrow r$. To tame the
divergence, we define regularized kernel
$$
 \langle r,\varphi|(H_s-\omega)^{-1}\\{\rm exp}(-tH_s^2)|r',\varphi'\rangle=
    \frac1{2\pi}\int\limits_{0}^{\infty}\frac{dp\,p\,{\rm exp}(-t\hbar^2v^2p^2)}{\hbar^2v^2p^2-\omega^2}\times
$$
\begin{equation}\label{a6}
   \times\sum\limits_{n\in \mathbb{Z}}e^{i\left(n+\frac s2\right)(\varphi-\varphi')}
   \left(
     \begin{array}{cc}
       a^{11}_{n,p}(r,\,r') & a^{21}_{n,p}(r,\,r')\\
                                                  a^{12}_{n,p}(r,\,r') & a^{22}_{n,p}(r,\,r')\\
     \end{array}
   \right),
\end{equation}
where $t>0$ is the regularization parameter. Now the limit
$r'\rightarrow r$ can be taken safely, yielding
$$
    {\rm tr} \langle r,\varphi|(H_s-\omega)^{-1}{\rm exp}\left(-tH_s^2\right)|r,\varphi\rangle=\frac
    \omega\pi\int\limits_{0}^{\infty}\frac{dp\,p\,{\rm exp}(-t\hbar^2v^2p^2)}{\hbar^2v^2p^2-\omega^2}=
$$
\begin{equation}\label{a7}
    =\frac{\omega}{2\pi\hbar^2v^2}E_1(-t\omega^2),
\end{equation}
where
$$
E_1(u)=\int\limits_{u}^{\infty}\frac{du}{u}e^{-u}
$$
is the exponential integral (see Ref. \cite{Abra}). Note that,
actually, we have reiterated the derivation for the case of the
planar graphene sheet ($\eta=0$): the only difference is that in the
latter case all factors of 2 (including those at the order of Bessel
functions) in the right hand sides of Eq.(A.5) are absent.

Since Eq.(A.9) is independent of $r$ and $\varphi$, the integration
over the surface yields a factor of its area:
\begin{equation}\label{a8}
    {\rm Tr}(H_s-\omega)^{-1}{\rm exp}\left(-tH_s^2\right)=\frac
    {S\omega}{2\pi\hbar^2v^2}E_1(-t\omega^2).
\end{equation}
The divergence of the last quantity in the limit $t\rightarrow0_+$
does not contribute to the density of states, Eq.(34). This is due
to a specific form of a discontinuity of the exponential integral at
negative real values of its argument, ${\rm Im}\,E_1(-u\mp i0)=\pm
i\pi\,\,(u>0)$. Consequently, we get finite result (35).

\renewcommand{\thesection}{B}
\renewcommand{\theequation}{\thesection.\arabic{equation}}
\setcounter{section}{1} \setcounter{equation}{0}

\section*{Appendix B}

The partial hamiltonian corresponding to $n=n_c$ takes form, see
Eqs.(45)-(47) and (52),
\begin{equation}\label{b1}
    h=\hbar v\left(
               \begin{array}{cc}
                 0 & \partial_r+r^{-1}(1-F) \\
                 -\partial_r-r^{-1}F & 0 \\
               \end{array}
             \right).
\end{equation}
Let $h$ be defined on the domain of functions $\xi^0(r)$ that are
regular at $r=0$. Then its adjoint $h^\dag$ which is defined by
relation
\begin{equation}\label{b2}
    \int\limits_{0}^{\infty}dr\,r(1-\eta)[h^\dag\xi(r)]^\dag\xi^0(r)=
\int\limits_{0}^{\infty}dr\,r(1-\eta)[\xi(r)]^\dag h\xi^0(r)
\end{equation}
acts on the domain of functions $\xi(r)$ that are not necessarily
regular at $r=0$. So the question is whether the domain of
definition of $h$ can be extended, resulting in both the operator
and its adjoint being defined on the same domain of functions. To
answer this, one has to construct the eigenspaces of $h^\dag$ with
complex eigenvalues. They are spanned by the linearly independent
square-integrable solutions corresponding to the pair of purely
imaginary eigenvalues,
\begin{equation}\label{b3}
    h^\dag\xi^\pm(r)=\pm iMv^2\xi^\pm(r),
\end{equation}
where $Mv^2$ is inserted for the dimension reasons. It is
straightforward to show that only one pair of such solutions exists
\begin{equation}\label{b4}
    \xi^\pm(r)=\frac 1N\left(
                         \begin{array}{c}
                           e^{\pm i\frac \pi 4}K_F(rMv/\hbar) \\
                           e^{\mp i\frac\pi4}K_{1-F}(rMv/h) \\
                         \end{array}
                       \right),
\end{equation}
where $N$ is a certain normalization factor. Thus, the deficiency
index is equal to (1,1), and, according to the Weyl-von Neumann
theory of self-adjoint operators (see Ref. \cite{Alb}), the
self-adjoint extension of operator $h$ is defined on the domain of
functions of the form
\begin{equation}\label{b5}
    \left(
      \begin{array}{c}
        f_{n_c,s}(r) \\
        g_{n_c,s}(r) \\
      \end{array}
    \right)=\xi^0(r)+c\left[\xi^+(r)-e^{-i\Theta}\xi^-(r)\right],
\end{equation}
where $c$ is a complex constant and $\Theta$ is a real continuous
parameter. Using the asymptotics of the Macdonald function at small
values of its argument (see Ref. \cite{Abra}), we get
\begin{equation}\label{b6}
\!\left(
      \begin{array}{c}
        f_{n_c,s}(r) \\
        g_{n_c,s}(r) \\
      \end{array}
    \right)\begin{array}{c}
             = \\[-0.2cm]
r\rightarrow 0
           \end{array}
\frac{2 c\,e^{-i\frac\Theta2}}{iN}\left(\!
                                \begin{array}{c}
                                  -\sin\left(\frac\Theta2+\frac\pi4\right)2^F\Gamma(F)(rMv/\hbar)^{-F} \\
                                  \cos\left(\frac\Theta2+\frac\pi4\right)2^{1-F}\Gamma(1-F)(rMv/\hbar)^{-1+F} \\
                                \end{array}\!
                              \right),
\end{equation}
which can be rewritten in the form of Eq.(53).

\newpage
\renewcommand{\thesection}{}
\renewcommand{\theequation}{\thesection.\arabic{equation}}
\setcounter{section}{1} \setcounter{equation}{0}

\newpage

\clearpage
\begin{figure}[t]
\begin{tabular}{l}
\includegraphics[width=120mm]{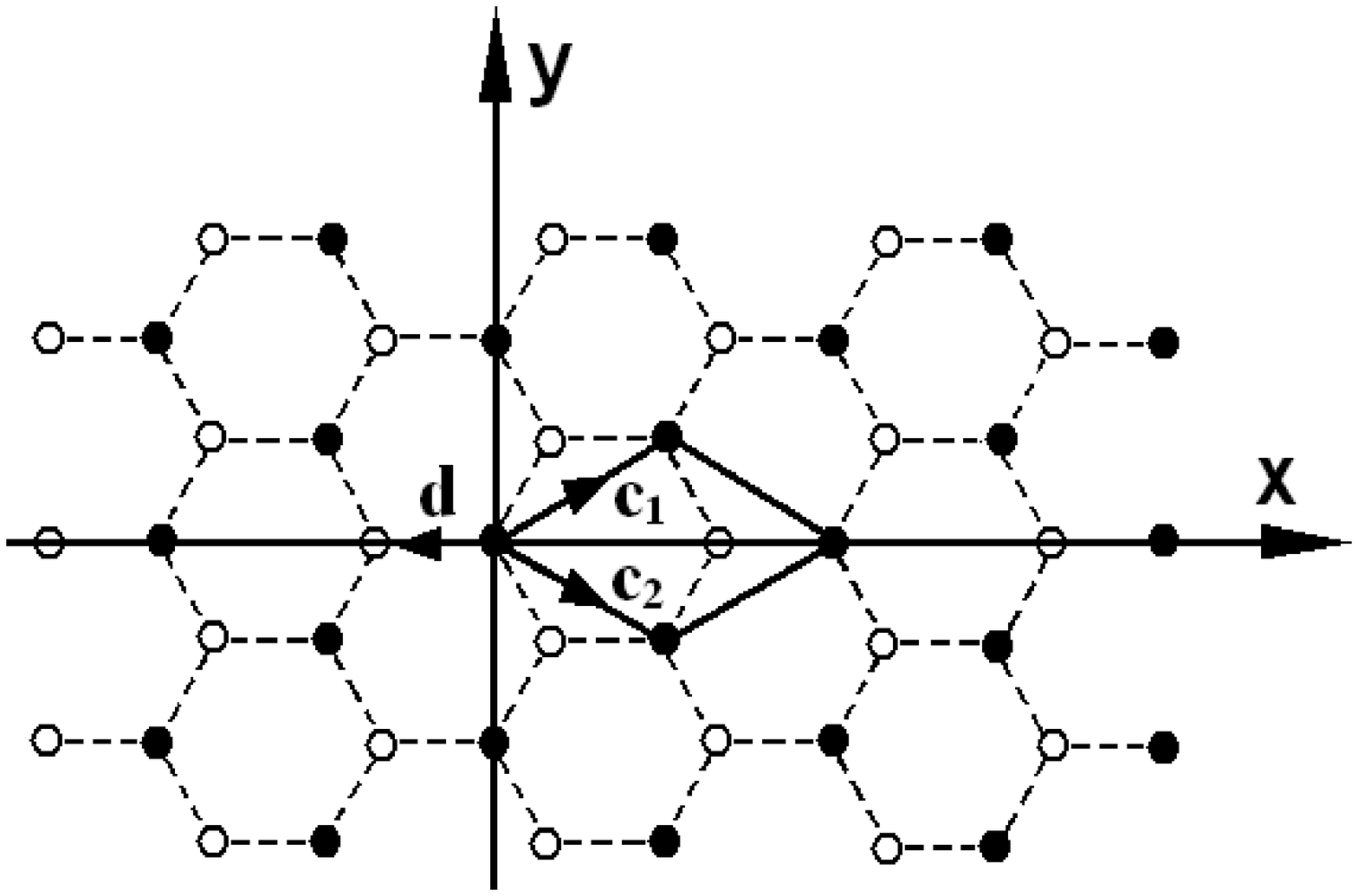}
\end{tabular}
\put(-125,-140){a)}
\end{figure}
\begin{figure}[h]
\begin{tabular}{l}
\includegraphics[width=120mm]{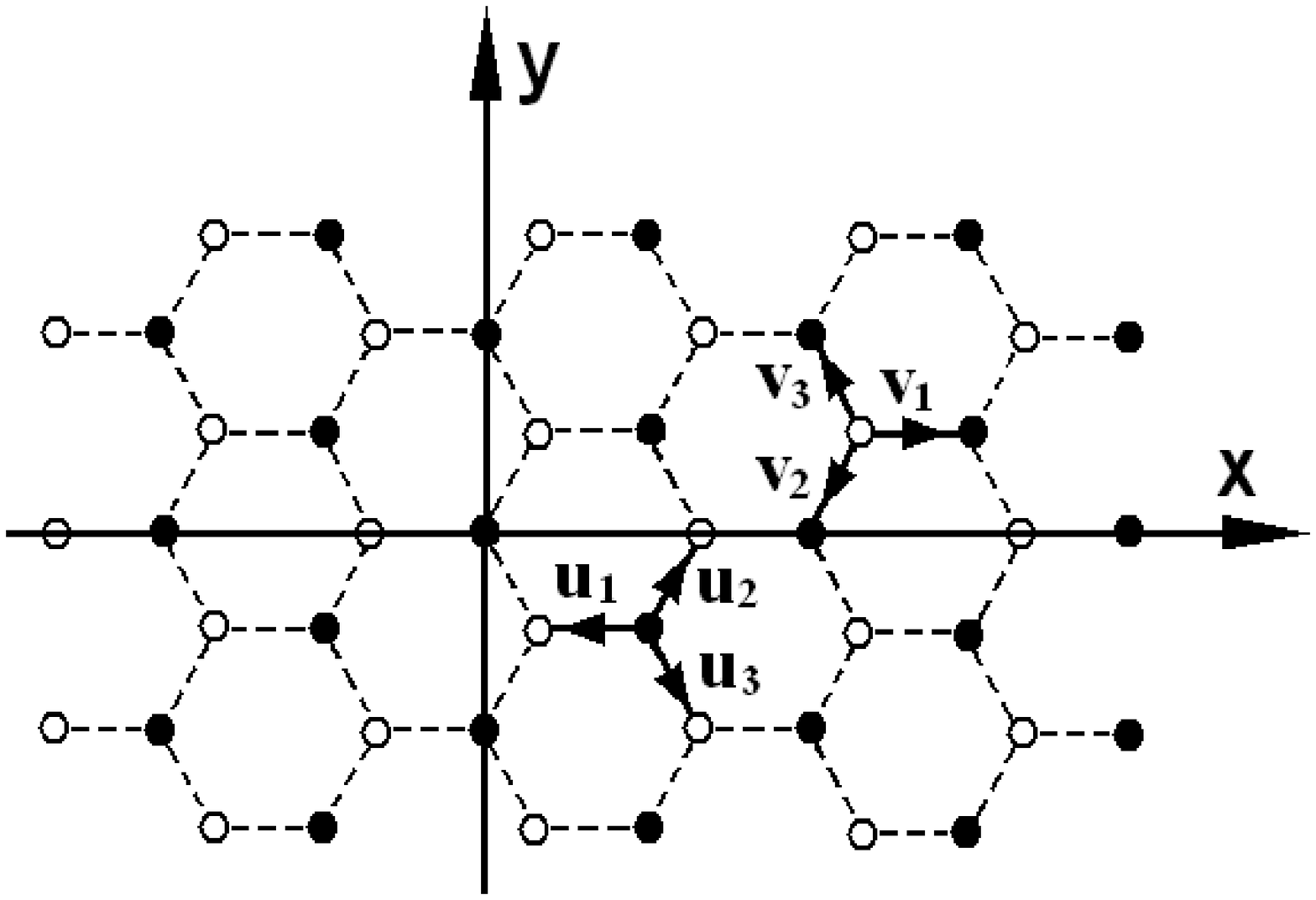}
\end{tabular}
\put(-125,-140){b)}
\caption{ The planar honeycomb lattice as a composition of two
triangular sublattices. The primitive cell with basis vectors ${\bf
c}_1$ and ${\bf c}_2$ is depicted in a), the triads ${\bf u}_j$ and
${\bf v}_j$ connecting different sublattices are depicted in b).}
\end{figure}

\clearpage
\begin{figure}[t]
\begin{tabular}{l}
\includegraphics[width=150mm]{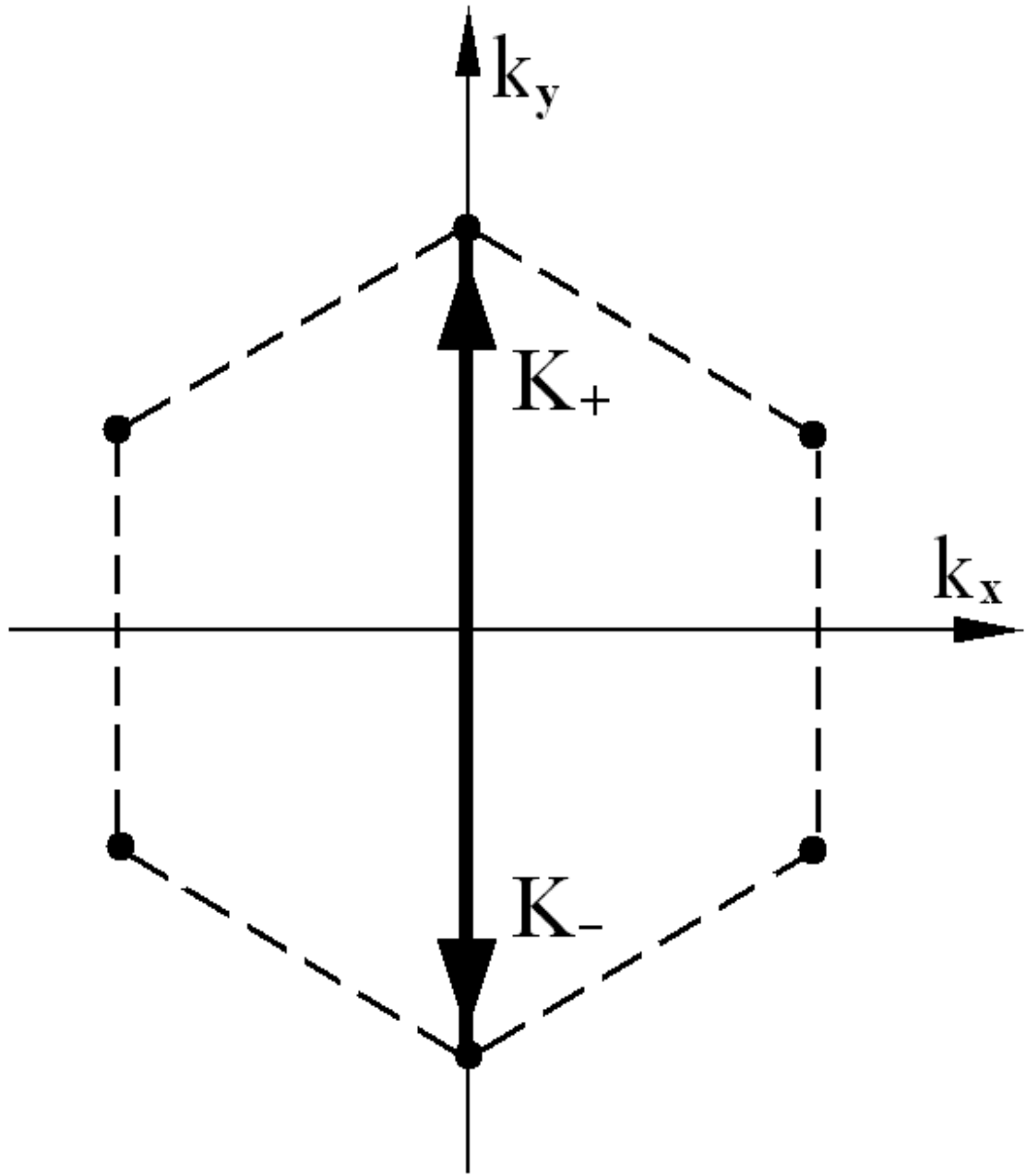}
\end{tabular}
\caption{ The first Brillouin zone is a hexagon with
opposite sides identified, and, therefore, next to neighbouring
corners are equivalent; two inequivalent ones can be chosen as lying
on a vertical line.}
\end{figure}

\clearpage
\begin{figure}[t]
\begin{tabular}{l}
\includegraphics[width=85mm]{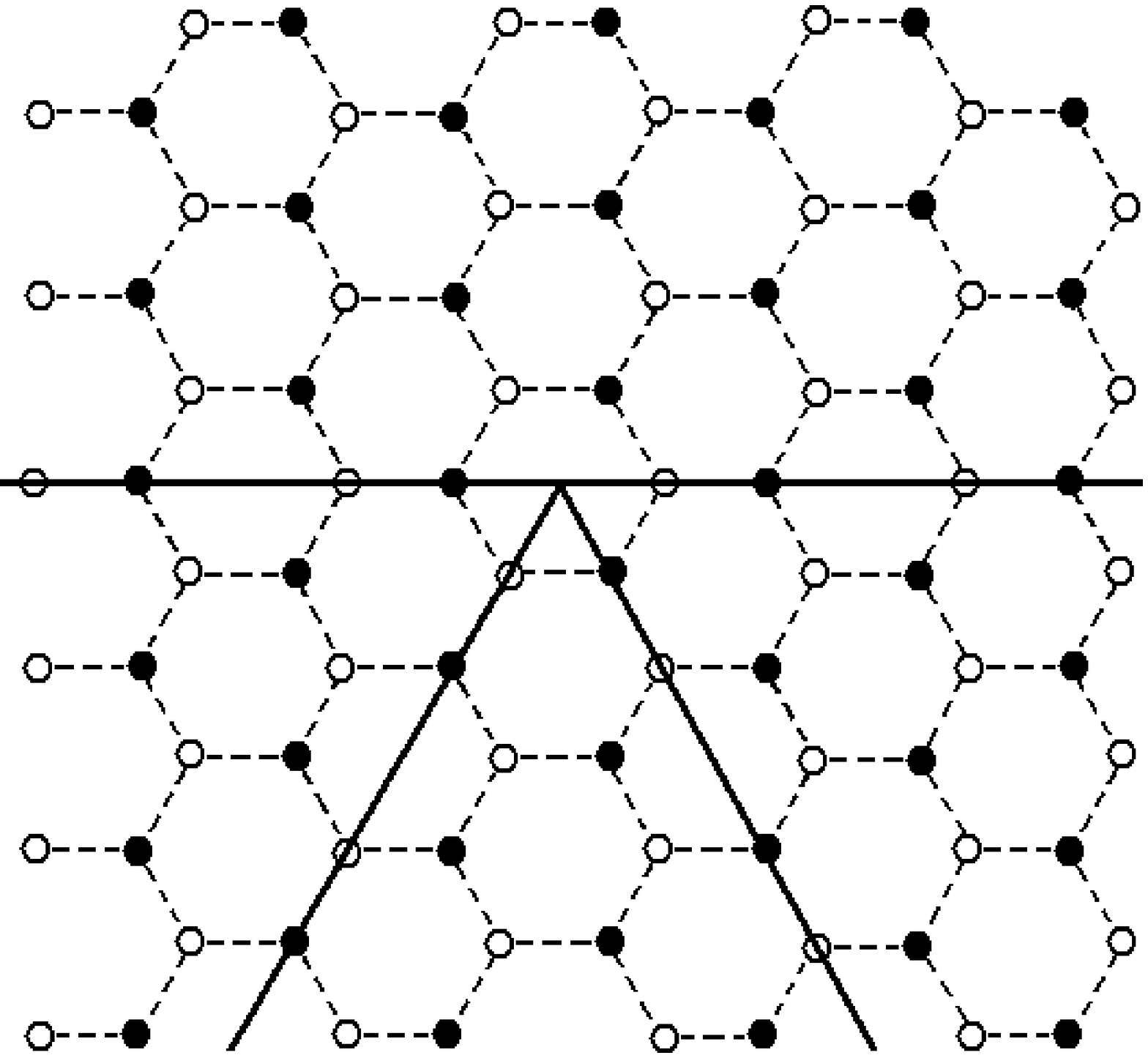}
\end{tabular}
\put(-70,-130){a)}
\end{figure}
\begin{figure}[h]
\begin{tabular}{l}
\includegraphics[width=85mm]{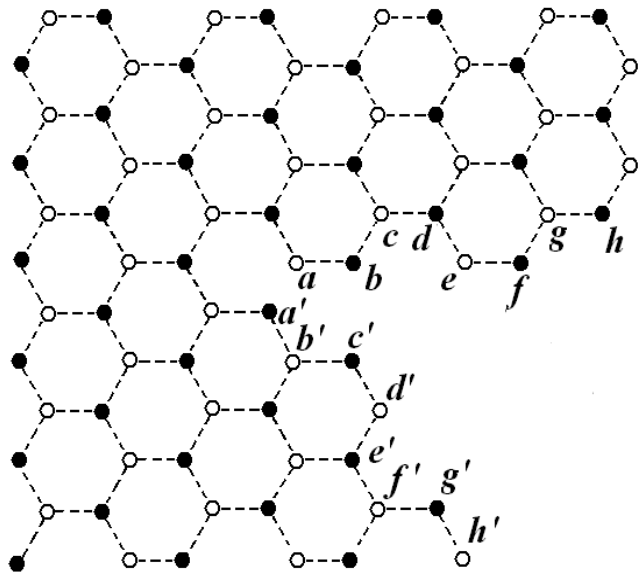}
\end{tabular}
\put(-70,-120){b)}
\caption{ Formation of a topological defect in graphene: a)
one, two, or three sectors of 60$^\circ$ are removed from the
lattice, b) if one sector is removed, then sites of different
sublattices are identified, $a$ and $a'$, $b$ and $b'$, $c$ and
$c'$, etc.}
\end{figure}

\clearpage
\begin{figure}[t]

\begin{tabular}{l}
\includegraphics[width=150mm]{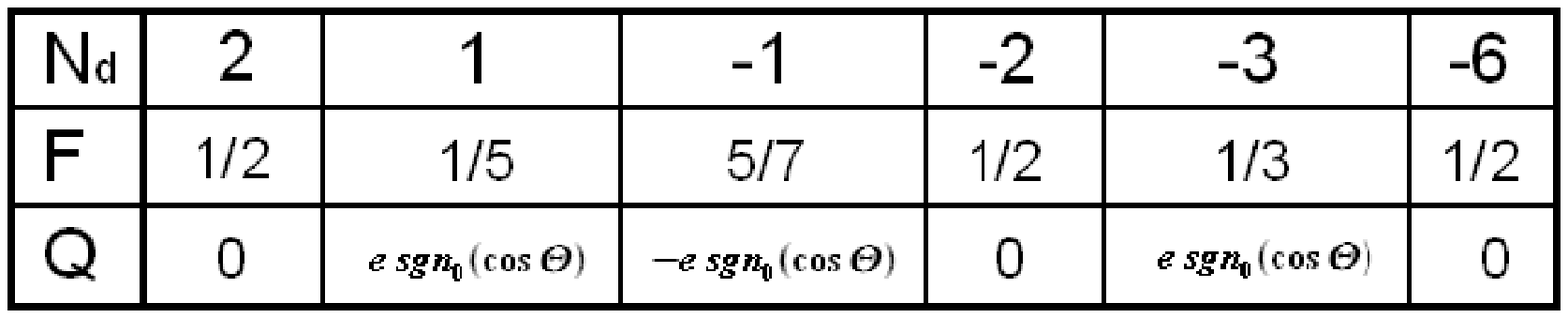}
\end{tabular}
\put(-240,130){Table:} \put(-410,115){The ground state charge in the
case of the existence of the one irregular } \put(-410,100) {mode in
the set of eigenmodes of the hamiltonian}
\end{figure}

\end{document}